\documentclass[a4paper,11pt]{article}
\usepackage{jinstpub}
\usepackage{graphicx}
\usepackage{siunitx}
\usepackage{booktabs}
\usepackage{hyperref}
\usepackage{lineno}
\usepackage{xspace}

\newcommand{\HeCF}{He:CF$_4$\xspace}
\newcommand{\TPC}{TPC\xspace}
\newcommand{\GEM}{GEM\xspace}
\newcommand{\GEMs}{GEMs\xspace}
\newcommand{\qCMOS}{qCMOS\xspace}

\title{The CYGNO experiment: a gaseous TPC with optical readout for rare events searches}

\author[1]{F.D. Amaro}
\author[2,3]{R. Antonietti}
\author[4,5]{E. Baracchini}
\author[6]{L. Benussi}
\author[6]{C. Capoccia}
\author[6,7]{M. Caponero}
\author[8]{L.G.M de Carvalho}
\author[9,10]{G. Cavoto}
\author[6]{I.A. Costa}
\author[6]{A. Croce}
\author[4,5]{M. D'Astolfo}
\author[10]{G. D'Imperio}
\author[6]{G. Dho}
\author[10]{E. Di Marco}
\author[1]{J.M.F. dos Santos}
\author[4,5]{D. Fiorina}
\author[10]{F. Iacoangeli}
\author[4,5]{Z. Islam}
\author[4,5]{H. P. Lima Jr}
\author[6]{G. Maccarrone}
\author[1]{R.D.P. Mano}
\author[4,5]{D. J. G. Marques}
\author[6]{G. Mazzitelli}
\author[2,3]{P. Meloni}
\author[9,10]{A. Messina}
\author[1]{C.M.B. Monteiro}
\author[8]{R.A. Nobrega}
\author[8]{I.F. Pains}
\author[6]{E. Paoletti}
\author[2,3]{F. Petrucci}
\author[4,5]{S. Piacentini}
\author[6]{D. Pierluigi}
\author[10]{D. Pinci}
\author[10]{F. Renga}
\author[6]{A. Russo}
\author[6,12]{G. Saviano}
\author[1]{P.A.O.C. Silva}
\author[13]{N. J. C. Spooner}
\author[6]{R. Tesauro}
\author[6]{S. Tomassini}
\author[4,5]{S. Torelli}
\author[9,10]{D. Tozzi}

\affiliation[1]{LIBPhys; Department of Physics; University of Coimbra; 3004-516 Coimbra; Portugal}
\affiliation[2]{Dipartimento di Matematica e Fisica; Universit\`a Roma TRE; 00146; Roma; Italy}
\affiliation[3]{Istituto Nazionale di Fisica Nucleare; Sezione di Roma Tre; 00146; Rome; Italy}
\affiliation[4]{Gran Sasso Science Institute; 67100; L'Aquila; Italy}
\affiliation[5]{Istituto Nazionale di Fisica Nucleare; Laboratori Nazionali del Gran Sasso; 67100; Assergi; Italy}
\affiliation[6]{Istituto Nazionale di Fisica Nucleare; Laboratori Nazionali di Frascati; 00044; Frascati; Italy}
\affiliation[7]{ENEA Centro Ricerche Frascati; 00044; Frascati; Italy}
\affiliation[8]{Universidade Federal de Juiz de Fora; Faculdade de Engenharia; 36036-900; Juiz de Fora; MG; Brasil}
\affiliation[9]{Dipartimento di Fisica; Universit\`a di Roma Sapienza; 00185; Roma; Italy}
\affiliation[10]{Istituto Nazionale di Fisica Nucleare; Sezione di Roma; 00185; Roma; Italy}
\affiliation[11]{Universidade Estadual de Campinas  - UNICAMP;  Campinas 13083-859; SP; Brazil}
\affiliation[12]{Dipartimento di Ingegneria Chimica; Materiali e Ambiente; Sapienza Universit\`a di Roma; 00185; Roma; Italy}
\affiliation[13]{Department of Physics and Astronomy; University of Sheffield; Sheffield; S3 7RH; UK}

\emailAdd{fabrizio.petrucci@uniroma3.it}

\abstract{\noindent
The CYGNO collaboration is developing a novel strategy for directional Dark Matter searches based on a gaseous Time Projection Chamber (TPC). The detector is optimized for the exploration of light (0.5–50 GeV) WIMPs-like particles and employs a He/CF$_{4}$ gas mixture at atmospheric pressure, sensitive to both spin-dependent and spin-independent interactions.
A key feature of the project is its optical readout, which relies on photon detection rather than charge collection.
In CYGNO detectors, electrons released by ionizing tracks drift toward an amplification stage of three Gas Electron Multipliers (GEMs). The electron avalanches generate scintillation light that is captured by scientific CMOS (sCMOS) cameras for high-resolution two-dimensional imaging and by Photomultiplier Tubes (PMTs) that provide a precise time profile along the drift direction. This allows a 3D event reconstruction, detailed energy deposition mapping, and effective topology and head-to-tail discrimination.
Building on the achievements of the 50 L prototype (LIME), which successfully operated underground at LNGS, the next step is the deployment of a 0.4 m³ demonstrator, CYGNO-04, to be completed in 2026. The demonstrator will validate scalability and confirm the advantages of the proposed technique.
Recent results from LIME highlight strong progress in 3D tracking and particle identification. The current status of CYGNO-04 and its role in advancing the program will be presented as well.}


\notoc

\begin{document}
\maketitle
\flushbottom

\newpage

\section{Introduction}\label{sec:intro}
The direct detection of Weakly Interacting Massive Particles-like (WIMPs-like) via elastic scattering off atomic nuclei remains one of the most promising approaches to search for dark matter (DM), in particular in the light (0.5–50~$GeV/c^2$) mass range. 
The expected signal has two characteristic signatures: an annual modulation of a few percent in the recoil rate due to the Earth's revolution around the Sun and an anisotropy in the distribution of the recoil angle due to the motion of the Solar System through the Milky Way in the apparent direction of the Cygnus constellation. The ability to measure the \emph{direction} of nuclear recoils (NR) provides a unique handle to distinguishing a DM signal from backgrounds and to enable the exploration within the so called neutrino fog.
The sensitivity to WIMPs-like particle with a low mass requires low detection thresholds and the capability to resolve short tracks with energies of few keV. A light target maximizes the energy transfer, while the use of a low density gas at ambient pressure extends NR ranges to the millimeter scale. This opens the possibility to exploit topological observables and even head--tail asymmetries. These considerations motivate a gaseous \TPC\ with a finely segmented optical readout with low noise.

\section{The CYGNO technique and the LIME prototype}\label{sec:technique}

\begin{figure}[b]
  \centering
    \includegraphics[scale=0.6]{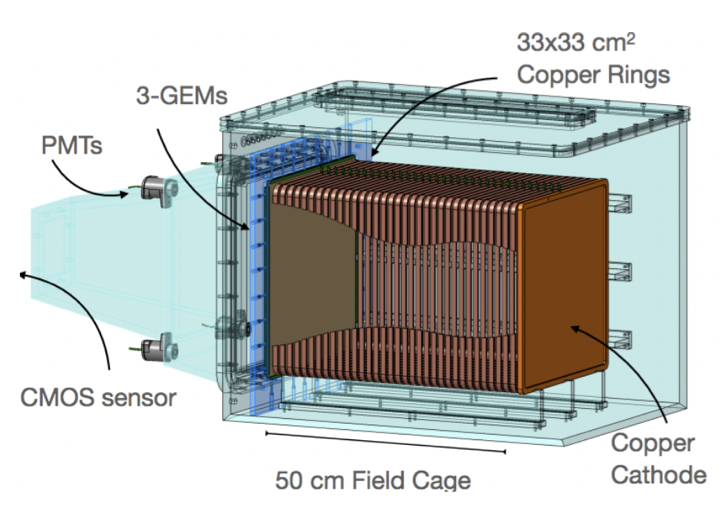}
    \includegraphics[scale=0.3]{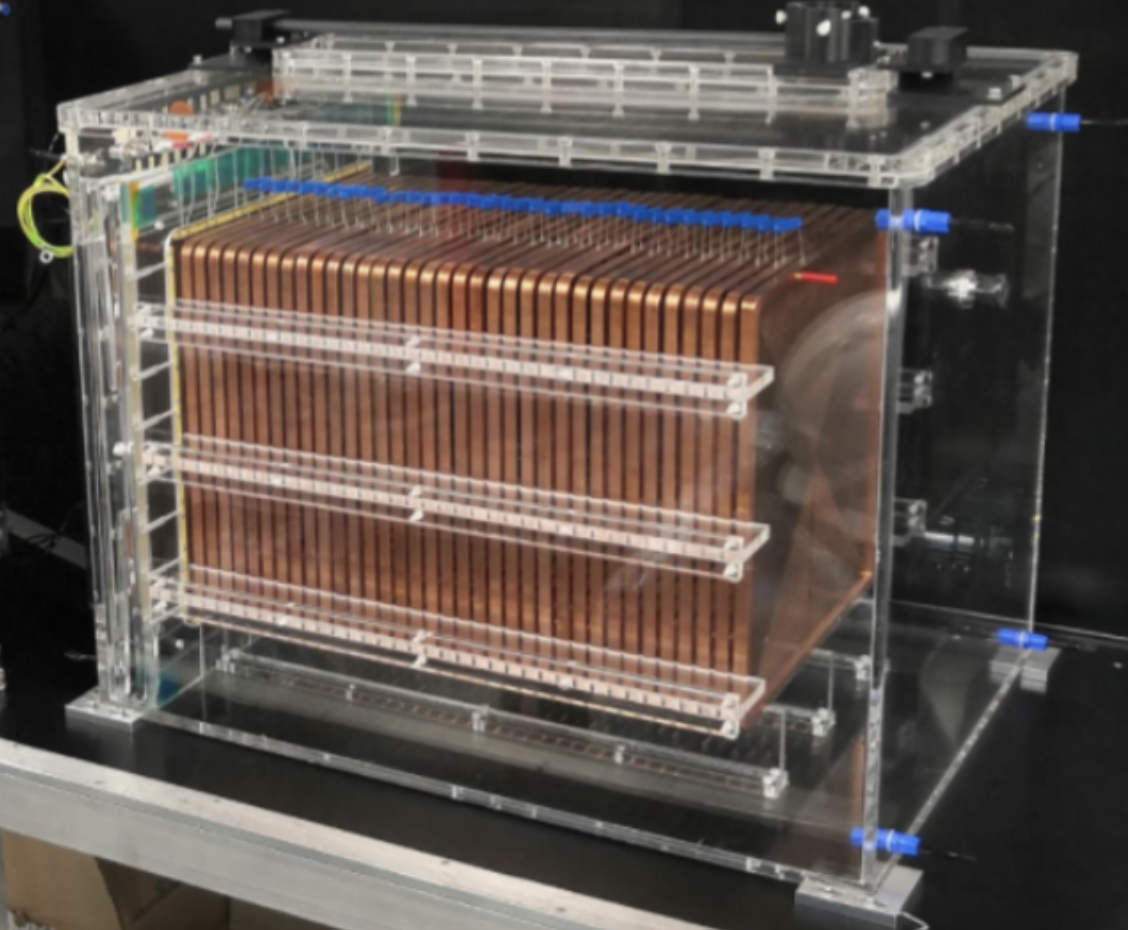}
  \caption{Schematics of the LIME prototype (left) explaining all the different components like the field cage, the amplification 3-\GEM\ stack, the gas volume and the readout optical system. The photo on the right shows the actual LIME apparatus.}
  \label{fig:technique}
\end{figure}

CYGNO \cite{Amaro2022Instruments} employs a \TPC\ filled with \HeCF\ (60:40) at ambient pressure. Ionization electrons, produced along the path of a charged particle, are drifted by a uniform electric field toward a 3-\GEM\ stack, where they undergo avalanche multiplication. In \HeCF, the avalanche produces scintillation light in the near-UV/visible band. The optical system exploits this light using a scientific CMOS (sCMOS) camera and fast PMTs.
This integrated system allows an high granularity imaging in the amplification (X--Y) plane with \(\mathcal{O}(10^2)\,\mu\mathrm{m}\) effective pitch, sufficient to resolve mm-scale tracks and their topology in the GEM plane, and 3D and inclination reconstruction exploiting the drift times as precisely measured by the PMTs. Moreover, the measurement of the track length, width, light density and head--tail asymmetries
can improve the discrimination power between electron and nuclear recoils. 

LIME \cite{Amaro2023EPJC} is the largest prototype built in the R$\&$D phase of the CYGNO project. 
As shown in Figure~\ref{fig:technique}, LIME is a 50~L active volume \TPC\ featuring thin (50~$\mu$m) \GEMs\ of area 33$\times$33~cm$^2$ and a drift gap of $\sim$50~cm. The optical readout uses a Hamamatsu ORCA Fusion sCMOS (2304$\times$2304~pixels, 6.5~$\mu$m pixel size projecting to 155~$\mu$m on the GEM plane, 2~counts/photon, 1~ph/pixel noise) and four PMTs placed along the readout plane diagonals. The prototype was commissioned overground at the Laboratori Nazionali di Frascati (LNF) of INFN to validate the gas system, the high-voltage distribution, the DAQ/trigger chain and data-handling, and was subsequently installed underground at at the Laboratori Nazionali del Gran Sasso (LNGS) of INFN for a test in a realistic environment. 

\begin{figure}[h]
 \centering
  \includegraphics[width=0.7\linewidth]{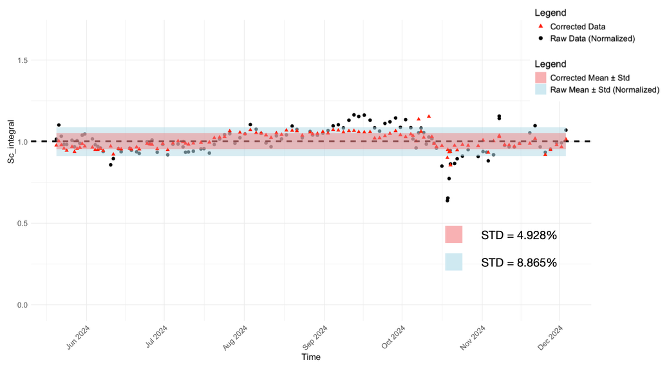}
  \caption{Stability of the measured light integral over time in a $\sim$6~months period.}
   \label{fig:calib}
\end{figure}

The underground data taking campaign lasted around 27 months and was divided in five periods with different shielding configurations. 
In Run~1 (no shielding), the external background model was validated; in Run~2 (4~cm Cu shielding) and
Run~3 (10~cm Cu shielding), the detector and shielding were commissioned, the internal backgrounds were studied and a data set with AmBe neutron source was acquired; in Run~4 (10~cm Cu + 40~cm water shielding), the water passive shielding for external neutron was added and the counting rate was further reduced; in Run~5 (back to 10~cm Cu shielding only) a high statistics and high stability set of runs were taken also intended for external neutron flux measurements. Overall, up to $\sim$1.2$\times10^7$ images were recorded, and trigger rates decreased from $\sim$34~Hz (no shielding) to $\sim$1~Hz (copper+water). The light yield measured with a collimated $^{55}$Fe source was stable at the 5\% level after corrections for gas-condition variations as shown in Figure~\ref{fig:calib}.

\section{Simulation framework}
A dedicated simulation framework was developed to model the full chain from energy deposition to sensor response \cite{simulation_paper}. The simulation includes: the primary ionization with realistic fluctuations; the diffusion and the absorption during the drift and across the \GEMs; the avalanche multiplication in a 3-\GEM\ cascade; the gain saturation caused by the charge density screening; the production of scintillation photons during the avalanche and their transport; the optical acceptance to the camera and the PMTs and a realistic description of the sensor noise. The model is able to closely reproduce the response linearity, the energy resolution and also the distributions of track-shape observables (length, width, light-density), supporting the reliability of the response model. 

\begin{figure}[htbp]
\begin{minipage}[b]{.47\textwidth}
  \centering
  \includegraphics[width=0.99\linewidth]{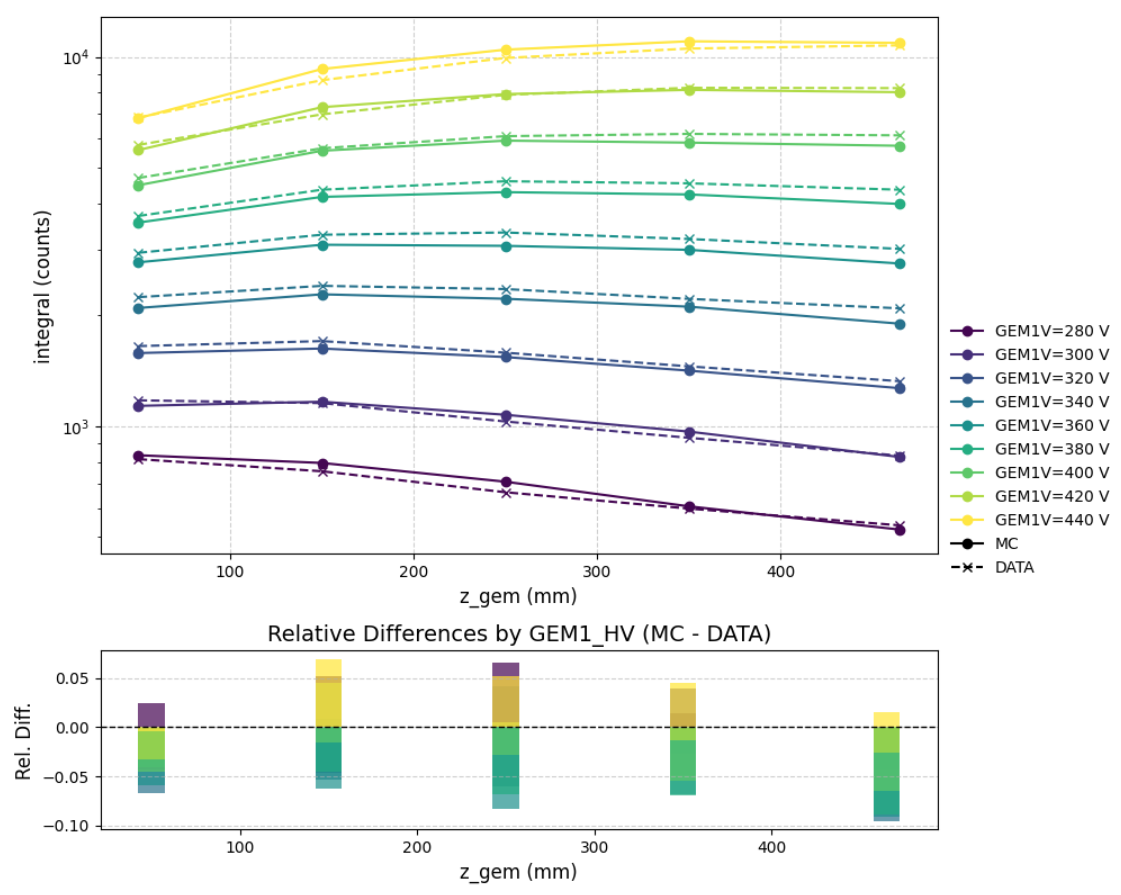}
      \caption{Comparison between data (dots and solid lines) and Monte Carlo simulation (crosses and dashed lines) for the average light integral as a function of the drift distance for different applied GEM voltages.}
  \label{fig:gem_sat}
  \end{minipage}
  \hfill
      \begin{minipage}[b]{.47\textwidth}
  \centering
  \includegraphics[width=0.9\linewidth]{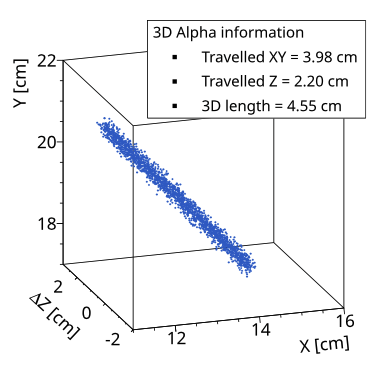}
      \caption{3D representation of the track of an alpha particle as reconstructed combining PMT and camera information \cite{Marques2025EPJC}.}
  \label{fig:3D_track}
  \end{minipage}
\end{figure}

The \GEM\ gain saturates at high local charge densities as ions backflow into the amplification region and partially screen the electric field. A simple model \cite{saturation_paper} was implemented in the simulation to scale the effective gain with the density of the avalanche, and tuned on data. As shown in figure~\ref{fig:gem_sat}, the model reproduces the average spot light yield as a function of drift distance and for different applied GEM voltages with better than 10\% accuracy over roughly two orders of magnitude in signal amplitude.

\section{Nuclear recoils and alpha-particle 3D reconstruction}\label{sec:nr}
To validate the detector response also to nuclear recoils, AmBe neutron source data were collected underground.  After calibration and normalization, the measured spectra are in reasonable agreement with the simulation. Work is ongoing to refine NR-specific reconstruction and discrimination algorithms, including $z$-dependent thresholds and position-dependent response corrections. 

Independently, the 3D reconstruction capability in the CYGNO approach was demonstrated starting with alpha particles \cite{Marques2025EPJC}. The method combines the X--Y image from the sCMOS with timing observables from the four PMTs (time-over-threshold and peak positions) to infer the drift component $\Delta z$ and the zenith angle. Cluster association across image and waveforms yields a complete 3D track definition; an example is shown for a candidate $\alpha$ particle in Figure~\ref{fig:3D_track}. The distribution of the reconstructed 3D track length exhibits the expected correlation with energy and reveals distinct peaks consistent with internal contaminants: a dominant \(^{222}\)Rn contribution and additional features attributable to the \(^{238}\)U and \(^{232}\)Th chains \cite{Marques2025Thesis}. The approach, validated on high SNR $\alpha$ tracks, is being extended to low energy NRs. 

\section{Preliminary evaluation of the sensitivity to a DM signal}\label{sec:sensitivity}
A preliminary DM sensitivity derived from a \(0.81\,\mathrm{kg\,day}\) LIME exposure was computed under the simplified assumption of a counting experiment. Results are already competitive with contemporary directional searches~\cite{Antonietti2025Thesis}. Analyses in progress incorporate realistic $z$-dependent thresholds and reconstruction performance, and will combine energy and angular information to maximize discovery and exclusion reach.

\section{The CYGNO-04 demonstrator}\label{sec:cygno04}
The next goal of our collaboration is to build a large volume demonstrator with the idea to demonstrate the full scalability of the technique in view of a large scale experiment. CYGNO-04 is a 0.4~m$^3$ \TPC\ with two drift volumes and a central cathode to be installed in LNGS Hall~F. The inner detector is hosted in a sealed PMMA vessel and surrounded by a passive shielding including 4~cm of radiopure copper (external vessel), 6~cm of copper from the OPERA experiment and $\sim$100~cm of water. The amplification stage uses V-bonded \GEMs\ to reduce light reflections and with random segmentation to reduce the dead areas between segments. Each side is optically read out by three \qCMOS\ cameras (ORCA QUEST2 class) and eight PMTs, enabling substantial gains in granularity and coverage relative to LIME. Civil works are complete; materials and services are procured; the inner detector and the vessel are expected to be ready in spring 2026 followed by a full detector commissioning.

\section*{Acknowledgments}
This project received funding from the European Union's Horizon~2020 research and innovation program through ERC Grant No.~818744.

\end{document}